\documentclass[fleqn,usenatbib]{mnras}

\usepackage[T1]{fontenc}

\DeclareRobustCommand{\VAN}[3]{#2}
\let\VANthebibliography\thebibliography
\def\thebibliography{\DeclareRobustCommand{\VAN}[3]{##3}\VANthebibliography}

\usepackage{graphicx}	
\usepackage{amsmath}	
\usepackage{amssymb}	
\usepackage{newtxtext,newtxmath}

\title[The expected kinematic matter dipole is robust against source evolution]{The expected kinematic matter dipole is robust against source evolution}

\author[von Hausegger]{
Sebastian {von Hausegger}$^{1}$\thanks{E-mail: sebastian.vonhausegger@physics.ox.ac.uk}
\\
$^{1}$Department of Physics, University of Oxford, Parks Road, Oxford OX1 3PU, United Kingdom}

\date{Accepted XXX. Received YYY; in original form ZZZ}

\pubyear{2024}

\begin{document}
\label{firstpage}
\pagerange{\pageref{firstpage}--\pageref{lastpage}}
\maketitle

\begin{abstract}
Recent measurements using catalogues of quasars and radio galaxies have shown that the dipole anisotropy in the large-scale distribution of matter is about twice as large as is expected in the standard $\Lambda$CDM model, indeed in any cosmology based on the Friedman-Lema\^itre-Robertson-Walker (FLRW) metric. This expectation is based on the kinematic interpretation of the dipole anisotropy of the cosmic microwave background,~i.e. as arising due to our local peculiar velocity. The effect of aberration and Doppler boosting on the projected number counts on the sky of cosmologically distant objects in a flux-limited catalogue can then be calculated and confronted with observations. This fundamental consistency test of FLRW models proposed by Ellis\&Baldwin in 1984 was revisited recently arguing that redshift evolution of the sources can significantly affect the expected matter dipole. In this note we demonstrate that the Ellis\&Baldwin test is in fact robust to such effects, hence the $>5\sigma$ dipole anomaly uncovered recently remains an outstanding challenge to the $\Lambda$CDM model.
\end{abstract}

\begin{keywords}
large-scale structure of Universe -- cosmology: observations -- cosmology: theory\vspace{-.7cm}
\end{keywords}



\section{Introduction}

The standard $\Lambda$CDM cosmological model describes a Universe in which small primordial inhomogeneities superimposed on a smooth background grow via gravitational instability over cosmic time. The background is defined by the FLRW metric of space-time that allows the evolution to be computed when the energy-momentum tensor is that of an ideal fluid. While this choice of metric was  motivated by simplicity rather than observational evidence, its implication  that the Universe should look homogeneous and isotropic to any comoving observer was promoted to the `Cosmological Principle' (CP) \citep{1933ZA......6....1M}. The validity of all conclusions that follow from interpreting data in the standard cosmological framework, including the inference of $\Lambda$ itself,  rests on whether the CP holds for the real Universe.

We are not in fact comoving observers; we infer this from the observation of a dipole anisotropy in the Cosmic Microwave Background (CMB) which is interpreted as due to our peculiar velocity $\beta\,=v/c\sim 10^{-3}$ with respect to the rest frame in which the CMB is isotropic~\citep{Stewart:1967ve,Peebles:1968zz}. This is called the `CMB frame', and the CP implies that the large-scale distribution of matter should also be isotropic in this frame. Hence in order to interpret cosmological data in the standard $\Lambda$CDM framework, observables measured in the heliocentric frame are boosted to the CMB frame. This has shown good  agreement with a wide range of data, moreover direct tests of homogeneity have so far supported the CP. For example counts-in-cells of quasars \citep[e.g.][]{Laurent:2016eqo} indicate homogeneity is reached on scales larger than a few hundred Mpc, although worryingly our local peculiar `bulk flow' shows no signs of dying out even on such scales \citep[e.g.][]{Watkins:2023rll}. With the exploration of cosmic structure on increasingly large scales, tests of the CP underlying $\Lambda$CDM have become increasingly important.\\

Long before it was robustly possible in practice, \citet{1984MNRAS.206..377E} (hereafter EB84) proposed a simple test of the CP via the dipole anisotropy in the projected number density of flux-limited catalogues of sources with power-law spectra, $S \propto \nu^{-\alpha}$, and integral source counts per unit solid angle, $\mathrm{d}N/\mathrm{d}\Omega(S>S_*) \propto S_*^{-x}$. A moving observer would see the source flux density to be Doppler boosted up or down, above or below the flux limit $S_*$, such that sources would move into or out of the sample, depending on their flux and their location on the sky with respect to the boost direction, while the usual aberration ($\propto \beta$) simultaneously displaces the sources towards the direction of motion. Consequently the observer moving with respect to the CMB frame would see a matter dipole aligned with the CMB dipole but with amplitude\vspace{-.05cm}
\begin{align}
    \tilde{\mathcal{D}}_{\rm kin} = \left[ 2 + \tilde x(1+\tilde\alpha) \right] \beta.
    \label{eq:D_EB}
\end{align}\vspace{-.05cm}\!\!
where, as we will elaborate below, $\tilde\alpha$ and $\tilde x$ refer to these quantities evaluated at the flux limit $S_*$.
The measurement of the matter dipole thus constitutes an independent measurement of the observer's velocity, that according to the CP must agree with the velocity inferred from the CMB dipole.

In the past two decades, with the onset of large cosmic surveys, the EB84 test has been performed with samples of high-redshift sources generally finding agreement in the direction of the matter dipole with that of the CMB, yet showing a tendency for the dipole amplitude to be 2--3 times larger than expected~\citep[e.g.][]{Blake:2002gx,Singal:2011dy,Gibelyou:2012ri,Rubart:2013tx}. The results, obtained using radio source catalogues, were however of marginal significance, because of low source counts as well as survey and methodological limitations, demonstrating the need for a larger, independent sample and a thorough study of systematic effects. \cite{Secrest:2020has} did the first such study, finding a matter dipole in mid-IR quasars which is twice as large as expected, rejecting the null hypothesis (Eq.~\ref{eq:D_EB}) with a significance of 4.9$\sigma$. This was later improved to a $>\!5\sigma$ result by \cite{Secrest:2022uvx} who included more quasars, as well as analysing an independent radio catalogue, and also confirmed by \cite{Dam:2022wwh} in a Bayesian reanalysis.\newpage

In parallel, theoretical studies were underway to check if the theoretical expectation for the matter dipole is affected by evolution of the catalogue's luminosity function~\citep[e.g.][]{Chen:2014bba,Maartens:2017qoa}. Could the discrepancy between the observed matter dipole amplitude and the EB84 prediction (Eq.~\ref{eq:D_EB}) be due to such effects? To provide an intuitive understanding of how redshift evolution of the sources and their properties can affect the dipole amplitude in the projected number counts, \citet{Dalang:2021ruy} (hereafter DB22) made a connection between the evolution of the luminosity function and the  EB84 parameters $\alpha$ and $x$ \citep[the latter is also called the `magnification bias', e.g.][]{Challinor:2011bk} by promoting them to be functions of comoving distance, $r$, viz. $\alpha(r)$ and $x(r)$. The expected dipole amplitude for a given $\beta$ now reads
\begin{align}
    \mathcal{D}_{\rm kin}=\int_0^\infty{\rm d}r\,f(r)\left[2+x(r)\left(1+\alpha(r)\right)\right]\beta,\label{eq:D_DB}
\end{align}
where $f(r)$ is the normalised distribution of sources in the catalogue to be defined below. DB22 proved that this expression is equivalent to the formulations that require knowledge of the sources' luminosity function. By  observing that Eq.~(\ref{eq:D_DB}) does not obviously reduce to Eq.~(\ref{eq:D_EB}), they cast doubt on the significance of the result obtained by \citet{Secrest:2020has,Secrest:2022uvx}, and called for the determination of the luminosity function and its evolution as the only viable approach to correctly predict the expected kinematic matter dipole amplitude.

In contrast, the seminal paper by EB84 stated clearly:
\begin{quote}
\textit{``The great power of this test
is that [...] the result must hold [...] irrespective of selection effects or \underline{source evolution}, as long as the forward and backward measurements are done in the identical manner.''}
\end{quote}
While clear, its rationale might not be obvious.  Here, we explicitly demonstrate that \citeauthor{1984MNRAS.206..377E}'s insight was indeed correct. We clarify the issue by first describing exactly how $\tilde x$ and $\tilde\alpha$ are determined in practice, and then show that, even given arbitrary source evolution, Eqs.~(\ref{eq:D_EB}) and~(\ref{eq:D_DB}) are perfectly equivalent.\vspace{-.2cm}

\section{Observed versus intrinsic quantities}

The argument of DB22, namely that Eq.~(\ref{eq:D_DB}) does not in general reduce to Eq.~(\ref{eq:D_EB}), starts by defining
\begin{align}
    x_{\rm eff} \equiv \int{\rm d}r\,f(r)x(r), &&
    \alpha_{\rm eff} \equiv \int{\rm d}r\,f(r)\alpha(r).\label{eq:eff}
\end{align}
It is then clear that $\mathcal{D}_{\rm eff} = \left[2+x_{\rm eff}\left(1+\alpha_{\rm eff}\right)\right]\beta$ only follows from Eq.~(\ref{eq:D_DB}) if $x(r)$ and $\alpha(r)$ are \emph{uncorrelated} over the domain of $f(r)$. We will show that care must be applied however when computing effective values for $x(r)$ and $\alpha(r)$. Specifically, $\tilde x$ is the index of a power-law approximation of the integral number counts ${\rm d}N/{\rm d}\Omega (S>S_*)$ \emph{at the threshold} $S_*$. Likewise $\tilde\alpha$ is the  average of only those sources that lie in the vicinity of $S_*$. These two requirements are not necessarily fulfilled by Eqs.~(\ref{eq:eff}). When they are fulfilled, we will see that $\tilde{\mathcal{D}}_{\rm kin}$~(\ref{eq:D_EB}) is indeed the same as $\mathcal{D}_{\rm kin}$ (\ref{eq:D_DB}).\vspace{-.3cm}

\subsection{The magnification bias $x$}

\begin{figure}
    \centering
    \includegraphics[width=0.99\columnwidth]{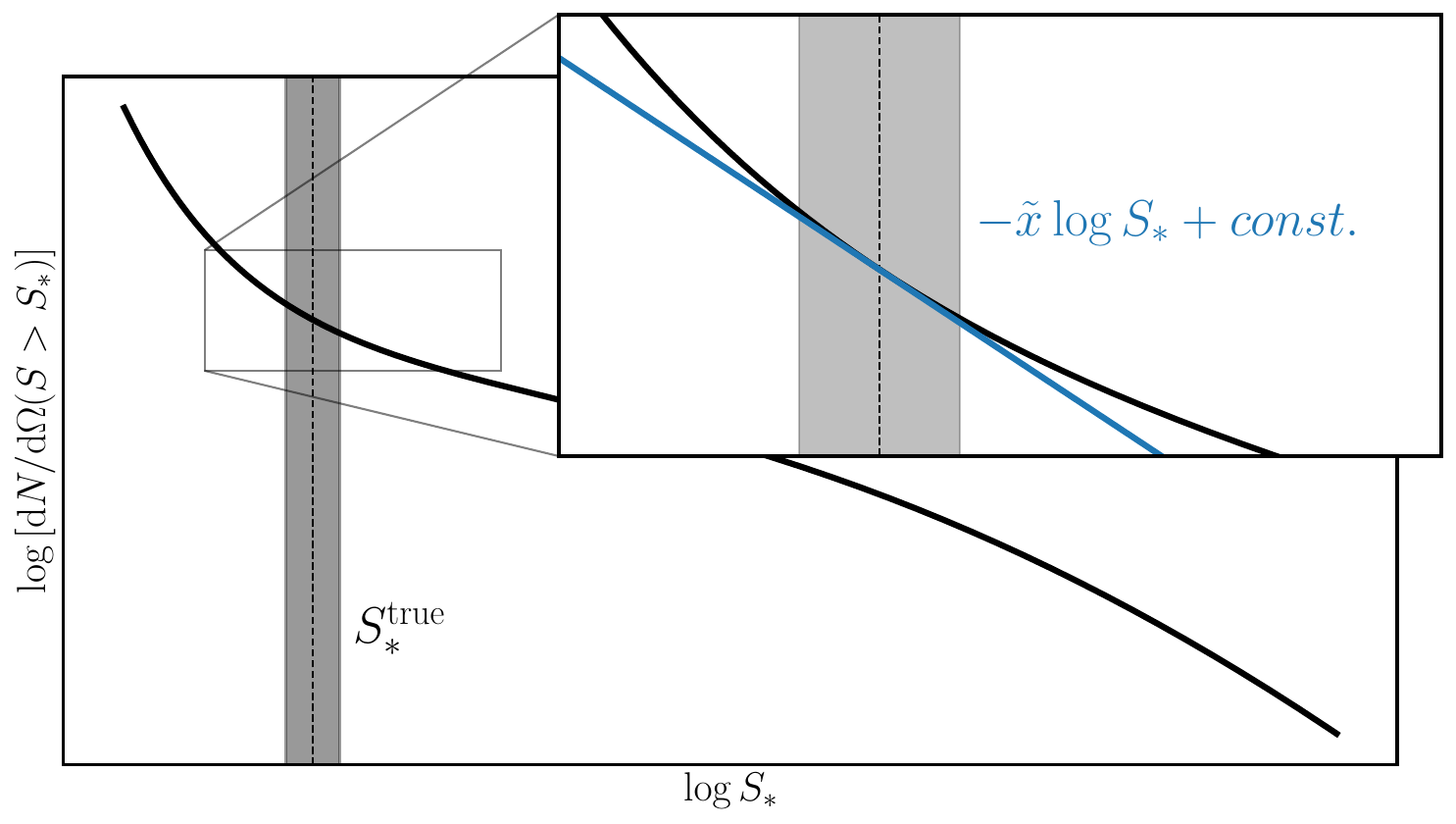}
    \caption{Sketch of the integral number counts (black curve) indicating that the index $\tilde x$ is to be measured at the limiting flux $S_*$ (dashed line) by fitting a power-law (blue line) in its neighbourhood. Only sources with flux density $S$ in a narrow (shaded) region of fractional width $\mathcal{O}(\beta)$ around $S_*$ (here chosen to be $S_*^{\rm true}$) contribute to the dipole amplitude, so the slope at higher or lower flux density is irrelevant.  In practice, measurement of the slope $\tilde x$ must be performed at $S_*^{\rm true}$ in bins small enough to resolve the shaded region.}\vspace{-10pt}
    \label{fig:xtilde}
\end{figure}
We follow the notation in DB22, but here, for completeness, state the general and implicit dependences $x(r)=x(r,S_*)$ and $f(r)=f(r,S>S_*)$.  In each shell of $r$ we find a distribution of sources with observed spectral flux density $S>S_*$,
\begin{align}
    \frac{{\rm d}N}{{\rm d}\Omega{\rm d}r}(r,S>S_*),\nonumber
\end{align}
that decreases with $S_*$ such that a power-law index $x$ is defined locally as
\begin{align}
\begin{split}
	x(r) \equiv& -\frac{\partial}{\partial\ln S_*}\left[\ln \frac{{\rm d}N}{{\rm d}\Omega{\rm d}r}(r,S>S_*) \right]\\
	=& -\frac{S_*}{\frac{{\rm d}N}{{\rm d}\Omega{\rm d}r}(r,S>S_*)}\frac{\partial}{\partial S_*}\left[ \frac{{\rm d}N}{{\rm d}\Omega{\rm d}r}(r,S>S_*) \right].\label{eq:xdef}
\end{split}
\end{align}
This is the magnification bias.  We also define the normalised distribution
\begin{align}
    f(r) \equiv \frac{{\rm d}N}{{\rm d}\Omega{\rm d}r}(r,S>S_*) \bigg/\int_0^\infty{\rm d}r\,\frac{{\rm d}N}{{\rm d}\Omega{\rm d}r}(r,S>S_*).\label{eq:f}
\end{align}
This distribution is not usually known for a given source sample and requires the knowledge of its luminosity function, rather than being directly observable.  In contrast, the flux distribution integrated over comoving distance, aka the integral source counts, \emph{is} an observed quantity
\begin{align}
    \frac{{\rm d}N}{{\rm d}\Omega}(S>S_*) = \int_0^\infty{\rm d}r\,\frac{{\rm d}N}{{\rm d}\Omega{\rm d}r}(r,S>S_*)
\end{align}
The EB84 test requires the determination of the integral source counts' slope \emph{only} at the flux density threshold $S_*$, as sources with flux densities far from it will not contribute to the dipole amplitude, e.g.~a source with very high $S\gg S_*$ will remain in the catalogue regardless of the boost applied.

As illustrated in Fig.~\ref{fig:xtilde}, one should thus measure $\tilde x$ \emph{locally} as the power-law approximation at the flux threshold, i.e.
\begin{align}
    \begin{split}
    \tilde x \equiv& -\frac{\partial}{\partial \ln S_*} \left[\ln \frac{{\rm d}N}{{\rm d}\Omega}(S>S_*)\right]\\
    =&-\frac{S_*}{\frac{{\rm d}N}{{\rm d}\Omega}(S>S_*)}\frac{\partial}{\partial S_*}\left[ \frac{{\rm d}N}{{\rm d}\Omega}(S>S_*) \right]\\
    =& -\frac{S_*}{\frac{{\rm d}N}{{\rm d}\Omega}(S>S_*)}\int_0^\infty{\rm d}r\,\frac{\partial}{\partial S_*}\left[ \frac{{\rm d}N}{{\rm d}\Omega{\rm d}r}(r,S>S_*) \right]\\
    =& \frac{1}{\int_0^\infty{\rm d}r\,\frac{{\rm d}N}{{\rm d}\Omega{\rm d}r}(r,S>S_*)}\int_0^\infty{\rm d}r\,\frac{{\rm d}N}{{\rm d}\Omega{\rm d}r}(r,S>S_*)x(r)\\
    =& \int{\rm d}r\,f(r)x(r) = x_{\rm eff},\label{eq:xtilde}
    \end{split}
\end{align}
where in the third line we used Eq.~\ref{eq:xdef}.  This shows that $x$ measured locally at $S_*$ in the observed integral number counts equals the appropriately weighted average as given by $x_{\rm eff}$.

\subsection{The spectral index $\alpha$}

While DB22 propose an effective value $\alpha_{\rm eff}$ as defined in Eq.~(\ref{eq:eff}) we reiterate that the relevant quantity should \emph{not} be averaged over all sources, but only over those that lie close enough to the threshold $S_*$ such that they actually contribute to the dipole amplitude.  In other words, since a source $i$ adds to the kinematic dipole only if its flux density $S_i$ lies within $S_*\pm(1+\alpha_i)\beta S_*$ (depending on its location on the sky), we must not also average over those sources' spectral indices $\alpha_i$ that are not included in this interval.  We therefore require a corresponding formalism that restricts the average value $\alpha_{\rm eff}$ of Eq.~(\ref{eq:eff}) which runs over all $S>S_*$ as per the definition of $f(r)$, to on that only includes sources with $S\sim S_*$.

To this effect, rather than integral source counts we require knowledge of differential source counts that can be evaluated at a given flux density $S$. The integral and differential source counts are related by
\begin{align}
    \frac{{\rm d}N}{{\rm d}\Omega{\rm d}r}(r,S>S_*) = \int_{S_*}^\infty{\rm d}S\,\frac{{\rm d}N}{{\rm d}\Omega{\rm d}r{\rm d}S}(r,S).\label{eq:dNdOdrdS}
\end{align}
This can be used to define a second normalised distribution function over comoving distance, $\tilde f(r) = \tilde f(r,S_*)$, only of sources with flux densities $S=S_*$,
\begin{align}
    \tilde f(r) \equiv& \left[\frac{{\rm d}N}{{\rm d}\Omega{\rm d}r{\rm d}S}(r,S) \middle/ \int_0^\infty{\rm d}r\, \frac{{\rm d}N}{{\rm d}\Omega{\rm d}r{\rm d}S}(r,S)\right]_{S=S_*}\nonumber\\
    =& \,\frac{{\rm d}}{{\rm d}S_*}\left[ \frac{{\rm d}N}{{\rm d}\Omega {\rm d}r}(r,S>S_*) \right] \bigg/ \int_0^\infty{\rm d}r\, \frac{{\rm d}}{{\rm d}S_*}\left[ \frac{{\rm d}N}{{\rm d}\Omega {\rm d}r}(r,S>S_*) \right]\nonumber\\
    =& \,x(r)\frac{{\rm d}N}{{\rm d}\Omega{\rm d}r}(r,S>S_*) \bigg/ \int_0^\infty{\rm d}r\, x(r)\frac{{\rm d}N}{{\rm d}\Omega{\rm d}r}(r,S>S_*),\label{eq:ftilde}
\end{align}
where we have used Eqs.~(\ref{eq:dNdOdrdS}) and again (\ref{eq:xdef}). Thus $\tilde\alpha$ is computed as the average spectral index of sources at the threshold $S_*$ as,
\begin{align}
    \tilde\alpha \equiv \int_0^\infty{\rm d}r\,\tilde f(r)\alpha(r)\label{eq:alphatilde},
\end{align}
which is generally different from $\alpha_{\rm eff}$~(\ref{eq:eff}).

\begin{figure}
    \centering
    \includegraphics[width=0.99\columnwidth]{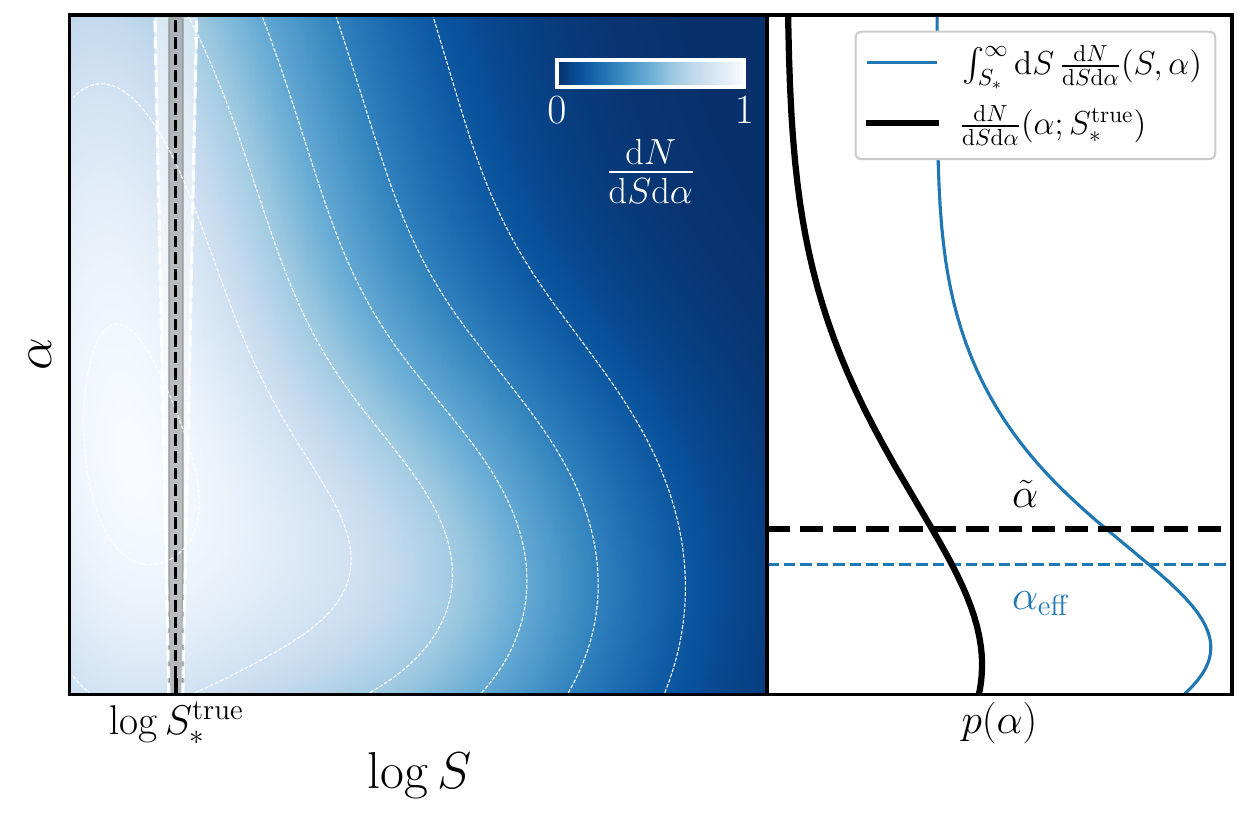}
    \caption{\textit{Left panel:} Illustrative sketch of the two-dimensional distribution of $S$ and $\alpha$.  The spectral index $\tilde\alpha$ is to be measured locally at $S_*$ (vertical dashed line).  As in Fig.~\ref{fig:xtilde}, only sources with flux density $S$ in a region (grey shaded) close to $S_*$ contribute to the dipole amplitude, so the $\alpha$ for other sources are irrelevant (The white shaded region indicates the true functional dependence on $\alpha$ of the relevant region, see main text for details).  The white dotted lines are for illustrative purpose only and denote iso-potential lines of the distribution.  \textit{Right panel:} Marginalised distributions of $\alpha$ when averaged over all values of $S>S_*$ (blue curve) or just over those in the vicinity of $S_*$ (black curve) leading to potentially different average values (horizontal dashed lines).} \vspace{-10pt}
    \label{fig:alphatilde}
\end{figure}

The difference between $\tilde\alpha$ and $\alpha_{\rm eff}$ is illustrated by the marginalised distributions shown  in Fig.~\ref{fig:alphatilde}: the 2-D distribution ${\rm d}N/{\rm d}S{\rm d}\alpha(S,\alpha)$ returns different 1-D distributions of $\alpha$, if averaged over all $S>S_*$, or only in the neighbourhood of $S_*$. This highlights the importance of correlations between $S$ and $\alpha$ in a given catalogue; these were in fact accounted for in the mock sky simulations of \cite{Secrest:2020has}.  While the area around the flux limit within which sources contribute to the dipole amplitude depends on $\alpha$ (as indicated by the white shaded region in the figure, and noted at the beginning of this subsection), in practice such corrections are negligible. Lastly, if $x(r)$ is a constant, then $\tilde\alpha=\alpha_{\rm eff}$ trivially.

\subsection{Dipole amplitude}

Using Eqs.~(\ref{eq:f}), (\ref{eq:xtilde}), (\ref{eq:ftilde}) and (\ref{eq:alphatilde}), $\tilde x$ and $\tilde\alpha$ can now be written as
\begin{align}
    \tilde x=&\!\!\int_0^\infty \!\!\! {\rm d}r\,\frac{{\rm d}N}{{\rm d}\Omega{\rm d}r}(r,S>S_*)x(r) \bigg/\!\!\!\int_0^\infty\!\!\! {\rm d}r\,\frac{{\rm d}N}{{\rm d}\Omega{\rm d}r}(r,S>S_*)
    \label{eq:xtilde2}\\
    \tilde \alpha=&\!\!\int_0^\infty\!\!\! {\rm d}r\,\frac{{\rm d}N}{{\rm d}\Omega{\rm d}r}(r,S>S_*)x(r)\alpha(r) \bigg/\!\!\!\int_0^\infty\!\!\! {\rm d}r\,\frac{{\rm d}N}{{\rm d}\Omega{\rm d}r}(r,S>S_*)x(r),
    \label{eq:alphatilde2}
\end{align}
using which we now compare Eq.~(\ref{eq:D_EB}) with Eq.~(\ref{eq:D_DB}). Due to the normalisation of $f(r)$ their first terms are trivially equal. Their second terms are equal according to  Eq.~(\ref{eq:xtilde}). The crucial term therefore is the last: 
\begin{align}
    \tilde x\tilde\alpha
    =&\int_0^\infty {\rm d}r\,\frac{{\rm d}N}{{\rm d}\Omega{\rm d}r}(r,S>S_*)x(r)\alpha(r) \bigg/ \int_0^\infty{\rm d}r\,\frac{{\rm d}N}{{\rm d}\Omega{\rm d}r}(r,S>S_*)\nonumber\\
    =& \int_0^\infty{\rm d}r\,f(r)x(r)\alpha(r),
\end{align}
since the numerator of Eq.~(\ref{eq:xtilde2}) cancels the denominator of Eq.~(\ref{eq:alphatilde2}). The dipole amplitude then reads
\begin{align}
    \begin{split}
    \tilde{\mathcal{D}}_{\rm kin} =& \left[2+\tilde x(1+\tilde\alpha)\right]\beta\\
    =& \int_0^\infty{\rm d}r\,f(r)\left[2+x(r)\left(1+\alpha(r)\right)\right]\beta = \mathcal{D}_{\rm kin}.\label{eq:D_measured}
    \end{split}
\end{align}
Therefore the dipole amplitude is determined entirely by the \emph{observed} quantities, $\tilde x$ and $\tilde\alpha$, which account implicitly for all source evolution effects. Moreover, as shown by DB22, in the absence of boundary terms as is appropriate for samples with smoothly decaying redshift distributions (see Discussion), there is exact equivalence between the dipole amplitude computed using the luminosity function~\citep{Maartens:2017qoa,Nadolny:2021hti,Dalang:2021ruy,Guandalin:2022tyl}, and that obtained by taking $x$ and $\alpha$ to be $r$-dependent \citep{Dalang:2021ruy}. Thus we have also demonstrated that the theoretically expected kinematic matter dipole amplitude does \emph{not} require any knowledge of the luminosity function.

\section{Discussion}

We have shown that the kinematic matter dipole can be robustly predicted using only \emph{observed} quantities, as per the EB84 prescription~(\ref{eq:D_EB}), which turned out to be fully equivalent to adding up the contributions (\ref{eq:D_DB}) from shells of comoving distance. Whereas DB22 highlighted the explicit connection between variations in the \emph{intrinsic} sample properties $x$ and $\alpha$, and the corresponding computation in a fully general relativistic setting \citep[e.g.][]{Challinor:2011bk}, according to their own calculation (see eq.~36), the equivalence we have shown also implies equivalence with the kinematic matter dipole amplitude calculated using the luminosity function. This makes the determination of the luminosity function redundant in this context, vindicating the assertion of EB84 that source evolution is irrelevant in the computation of the kinematic matter dipole.

The difficulty in determining the luminosity function was nevertheless emphasised as a major ``theoretical systematic''  in predicting the kinematic matter dipole (\citet{Guandalin:2022tyl}, see also \citet{Maartens:2017qoa}). This was first illustrated by DB22 \citep[see also][]{Dalang:2021ruy:Corrections} who questioned the significance of the anomaly found by \citet{Secrest:2020has} by showing that a range of dipole amplitudes is predicted depending on the choice of the functional form of the  magnification bias $x(z)$ and evolution bias $b_\mathrm{e}(z)$. This uncertainty only arises due to not knowing the exact functional forms that derive directly from the (unknown) luminosity function of the source sample. However as we have shown, none of this is relevant here. Such uncertainties are entirely bypassed in ensuring to use solely the observed quantities $\tilde x$ and $\tilde\alpha$ in the EB84 test, as is both appropriate and adequate in the present context. Nevertheless the relation we have found between $\tilde x$ and $x(r)$ may prove useful in constraining the magnification bias via measurements of the integral source counts, without reference to the (uncertain) luminosity function as has otherwise been demonstrated by e.g.~\citet{Wang:2020ibf}.

The correlations first suspected between $x(r)$ and $\alpha(r)$ by DB22 are based on a definition of $\alpha$, namely $\alpha_{\rm eff}$, that does not feature in the EB84 test. These correlations have been investigated empirically by \citet{Dam:2022wwh} who invoked the principle of maximal entropy to bound their maximum effect to be below 17\% in the quasar sample studied by \cite{Secrest:2020has}.  As we have seen that such correlations need not be studied at all in order to obtain a robust estimate of the dipole amplitude defined in terms of~$\tilde\alpha$.

Instead, the possible correlation between $\alpha$ and $S$ at the flux limit $S_*$ \emph{does} play a role in the EB84 test, as can be seen by studying $\tilde\alpha$ in Fig.~\ref{fig:alphatilde}.  This provides justification for the procedure employed by \cite{Secrest:2020has} to simulate mock skies in order to evaluate the significance with which the dipole measurement rejects the null hypothesis (\ref{eq:D_EB}). First, $N$ points were distributed randomly on the sky in line with the assumption of isotropy. Each point then was randomly assigned a pair of values $S$ and $\alpha$, sampled from the empirical distribution function ${\rm d}N/{\rm d}S{\rm d}\alpha$. This respects possible correlations between $S$ and $\alpha$ in the data; in the above context, it amounts to ensuring that in each of the simulations, $\tilde\alpha$ is conserved. Later, \cite{Secrest:2022uvx} found that the results are nearly unchanged even when $S$ and $\alpha$ are picked \emph{independently}, which they  interpreted as implying that there is no significant source evolution. In light of the present work, we can now confirm that this intuition was correct.

In summary, for samples whose redshift distributions fall sufficiently smoothly towards both low and high redshifts, \emph{no} alterations to the EB84 dipole prediction are needed to take into account source evolution effects, as long as the power-law assumptions for the integral source counts and for the source spectra hold at the flux limit $S_*$. (This had not been necessary to state explicitly by EB84 who considered perfect power-laws throughout.)  Possible curvature in integral source counts ${\rm d}N/{\rm d}\Omega(S>S_*)$ may warrant small corrections to $\tilde x$, as illustrated by the difference between the blue and the black curves in Fig.~\ref{fig:xtilde}. Also the domain over which to integrate $\alpha$, illustrated by the white shaded area in Fig.~\ref{fig:alphatilde}, offers scope for such corrections. Both are expected to be small corrections to an already small effect of $\mathcal{O}(\beta)$, but can nonetheless be estimated \emph{directly} from data. Related corrections were proposed by \citet{Tiwari:2013vff} and implemented by \citet{Siewert:2020krp}. While they did not recognise the adequacy of performing the fit of $\tilde x$ only locally---e.g. ~\citet{Siewert:2020krp} say their fit ``extends for each survey over one decade''---they demonstrated the feasibility of the procedure, albeit with little effect on their conclusions.

Looking ahead, this work showed that the expected kinematic matter dipole amplitude $\mathcal{D}_{\rm kin}$ in general is a function of the flux threshold $S_*$, since	 $\tilde x$ and $\tilde \alpha$ themselves depend on $S_*$, opening up much ground for study (e.g., \citet{Murray:2021frz} used this observation to constrain the lensing dipole contribution). Most importantly however, the present formalism allows to make practical connections between theoretical calculations of the \emph{redshift-dependent} kinematic matter dipole and the one derived from observed quantities. While measuring the matter dipole's redshift dependence naturally requires the careful definition of redshift-bins in data, the theoretical computation of its kinematic expectation necessitates equally careful treatment of boundary terms, especially for narrow redshift bins with sharp cut-offs. These terms naturally vanish in samples as those considered here and in DB22~\citep[see also][]{Domenech:2022mvt}. Nevertheless, future extension of the present work to accommodate redshift-tomographic measurements of the matter dipole is key to building a full understanding of the matter dipole anomaly. Given that correspondingly suitable galaxy catalogues from current surveys are in reach this could not be more timely.

\section*{Acknowledgements}

It is a pleasure to thank Charles Dalang, Harry Desmond, Nathan Secrest, and especially Subir Sarkar, for useful discussions on this matter, and for comments on an initial draft.  I am also grateful for many helpful remarks by and exchanges with Camille Bonvin, Chris Clarkson, Ruth Durrer, Pedro Ferreira, and Roy Maartens following the first circulation of this manuscript.  Finally, I thank the referee for their constructive remarks and suggestions to improve the paper.

This work made use of the following python packages:
\texttt{numpy}~\citep{harris2020array}, \texttt{scipy}~\citep{2020SciPy-NMeth}, and \texttt{matplotlib}~\citep{Hunter:2007}.

For the purpose of open access, the author has applied a Creative Commons Attribution (CC BY) licence to any Author Accepted Manuscript version arising.

\section*{Data Availability}

No new data were generated or analysed in support of this research.



\bibliographystyle{mnras}
\bibliography{References} 






\bsp	
\label{lastpage}
\end{document}